# Dual Photonics Probing of Nano- to Submicron-Scale Structural Alterations in Human Brain Tissues/Cells and Chromatin/DNA with the Progression of Alzheimer's Disease

Fatemah Alharthi [1], Ishmael Apachigawo [1], Dhruvil Solanki [1], Sazzad Khan [2], Himanshi Singh [2], Mohammad Moshahid Khan [2,*] and Prabhakar Pradhan [1,*]

[1] Department of Physics and Astronomy, Mississippi State University, Mississippi State, MS 39762, USA;
[2] Department of Neurology, The University of Tennessee Health Science Center, Memphis, TN 38163, USA;

**Abstract:** Understanding alterations in structural disorders in tissue/cells/building blocks, such as DNA/chromatin in the human brain, at the nano to submicron level provides us with efficient biomarkers for Alzheimer's detection. Here, we report a dual photonics technique to detect nano- to submicron-scale alterations in brain tissues/cells and DNA/chromatin due to the early to late progression of Alzheimer's disease in humans. Using a recently developed mesoscopic light transport technique, fine-focused nano-sensitive partial wave spectroscopy (PWS), we measure the degree of structural disorder in tissues. Furthermore, the chemical-specific inverse participation ratio technique (IPR) was used to measure the DNA/chromatin structural alterations. The results of the PWS and IPR experiments showed a significant increase in the degree of structural disorder at the nano to submicron scale at different stages of AD relative to their controls for both the tissue/cell and DNA cellular levels. The increase in the structural disorder in cells/tissues and DNA/chromatin in the nuclei can be attributed to higher mass density fluctuations in the tissue and DNA/chromatin damage in the nuclei caused by the rearrangements of macromolecules due to the deposition of the amyloid beta protein and damage in DNA/chromatin with the progress of AD.

**Keywords:** light scattering; Alzheimer's disease; disorder strength; partial wave spectroscopy; confocal imaging; inverse participation ratio technique; mesoscopic physics

## 1. Introduction

According to several recent studies, cells and tissues are elastic scatterers of light [1–6]. This scattered light, which probes the cells/tissues, provides valuable information about the structural composition and arrangement of macromolecules in the cells due to the different scattering properties, such as the refractive index [7–9]. Several studies have shown light probing properties of cell/tissue disorders, even quantifying such disorders [2,10–19]. Based on the literature [20], we understand how light probing of samples in different stages of cancer provides valuable information for early detection and mitigation measures to understand the fundamental physical properties of the progression of diseases. Similarly, brain disorders such as Alzheimer's disease (AD) impact tissues and cells' structural characteristics at all length scales, ranging from microns to sub-microns. From the literature, we can ascertain that with the help of the mesoscopic physics analysis technique, cell abnormalities in disease conditions can provide appropriate structural biomarkers for disease detection [21,22].

The technique used in this paper, as described elsewhere [23,24], is known as partial wave spectroscopy (PWS). It uses backscattered signals from weakly disordered media to measure and statistically quantify the intensities of these signals due to changes in the nanoscale refractive index fluctuations $(dn)$ of cells/tissues [14,25–27]. The spatial density fluctuations $(d\rho)$ that occur due to the rearrangement of macromolecules in subcellular structures like DNA/chromatin allow us to use a single parameter to characterize the strength of disorder, $L_{d\text{-}PWS}$ [20,28,29]. Several results underline this technique's importance, which has been used to explore many structural disorder cases studied for disease detection and detailed in previous studies described elsewhere [23,24,30–33]. The methods discuss the mathematical and theoretical description of the PWS technique. In addition to this technique, our study requires knowledge of the chemical spatial density alterations in specific DNA/chromatin; this is obtained using the inverse participation ratio (IPR) technique using confocal imaging. This molecular-specific localization technique helped us to quantify the abnormalities in cellular structures in the brain caused by conditions such as stress and alcoholism [34–39].

Alzheimer's disease (AD), considered to be one of the deadliest diseases, is a neurodegenerative disease that affects a patient's cognitive abilities and causes memory loss [40–45]. The symptoms worsen as the disease progresses with time. AD has been ranked among the top seven deadliest illnesses in the United States by the Centers for Disease Control and Prevention, surpassing breast and prostate cancers [46]. Primarily, older people who are above age 65 are diagnosed with AD, but younger people can also be affected. It has been shown that structural changes in the brain at the cellular level start decades before the symptoms appear because of intracellular structural alterations at the nano to submicron scale before any clinical diagnosis can be made at the macro scale. Early abnormal accumulation of proteins that form amyloid plaques and neurofibrillary tangles (NFT) is thought to affect cognitive abilities and result in the development of other symptoms [47]. Since structural changes occur before any symptoms appear, detecting those abnormalities at a nanoscale level is necessary for early diagnosis [48].

Understanding and detecting any disorder from the early stages is paramount, even before it becomes apparent. Disorder (cancers/tumors) formation begins at the subcellular level, resulting in cell size changes and rearrangement of the basic building blocks of organisms [49–51]. However, these changes are undetectable due to limitations in conventional microscopy resolution (<200 nm). Histological studies have shown that the microscopic tissues become more damaged with the progression of terminal diseases like AD at the last stages. Studies on the structural changes in brain cells/tissues are limited. Hence, it is essential to study these progressions at their early stages using an appropriate structural probing technique that is capable of identifying changes at the submicron scale.

Using the dual photonic technique, this work analyzes three stages of AD in human brain cells/tissues. We probe these tissues using light/photonic scattering probes to obtain insight into the spatial and chemical density fluctuations from the backscattered signal and to detect and define AD stages using the structural disorder strength as our quantification parameter. As mentioned, there are little to no observable physical symptoms in these lethal diseases from the early onset, but at progressively later stages of the disease, structural symptoms become visible using microscopy, MRI, and CT scans. We are interested in statistically quantifying the subcellular changes in the cells/tissues of AD brain samples and comparing them to control/healthy brain samples in a specific region in the brain. We draw parallels and distinguish between physical features like $dn$ changes and $d\rho$ changes by quantifying the elastically scattered light intensity, and based on this, we can characterize the various stages of the disease. Using the PWS technique, we can use the structural disorder strength ($L_{d\text{-}PWS}$) as a potential biomarker [23]. To enhance our study, the IPR approach is utilized to further analyze the spatial structural disorder in the spatial DNA molecular mass density spatial fluctuations ($d\rho_{ms}$) of nuclear DNA components using confocal imaging ($L_{d\text{-}IPR}$) [35]. We additionally correlate these findings with DNA damage/structural rearrangements and Aβ pathology observed in AD patients. The following section discusses the PWS analysis of tissue samples, the IPR analysis of DNA from cell nuclei, supporting histopathology of DNA damage, and the results and applications.

## 2. Results

### 2.1. PWS Analysis of Thin Brain Tissues

We used a fine-focused PWS technique to analyze human brain tissues to measure ($L_{d\text{-}PWS}$) and compare different stages of AD. We compared these to understand how the cellular structure changes at the nano to submicron scales as the disease progresses. Our results revealed a significant increase in structural alterations at the cellular level with the progression of AD. To obtain ($L_{d\text{-}PWS}$), we need to determine the variance in $dn$ and its $l_c$ (as shown in Equations (5) and (6)). As discussed in the experimental section, we acquired the $dn^2 \times l_c$ from each pixel of the spectral images captured using CCD to compute the $L_{d\text{-}PWS}$ for each pixel. This is shown in a 2D color map with colors ranging from blue to red, indicating changes in $L_{d\text{-}PWS}$. Figure 1a–d show brightfield images of the human brain samples, with control (C), low AD (LAD), intermediate AD (IAD), and severe AD (SAD) collected using CCD. Figure 1a'–d' represent their corresponding $L_{d\text{-}PWS}$ images. It can be seen clearly in the figure of the $L_{d\text{-}PWS}$ map that in severe AD cases, there are substantial numbers of yellow and red spots in the color map, reflecting more fluctuations in the refractive index and thus an increase in the mass density fluctuations/rearrangements at the cellular/tissue level.

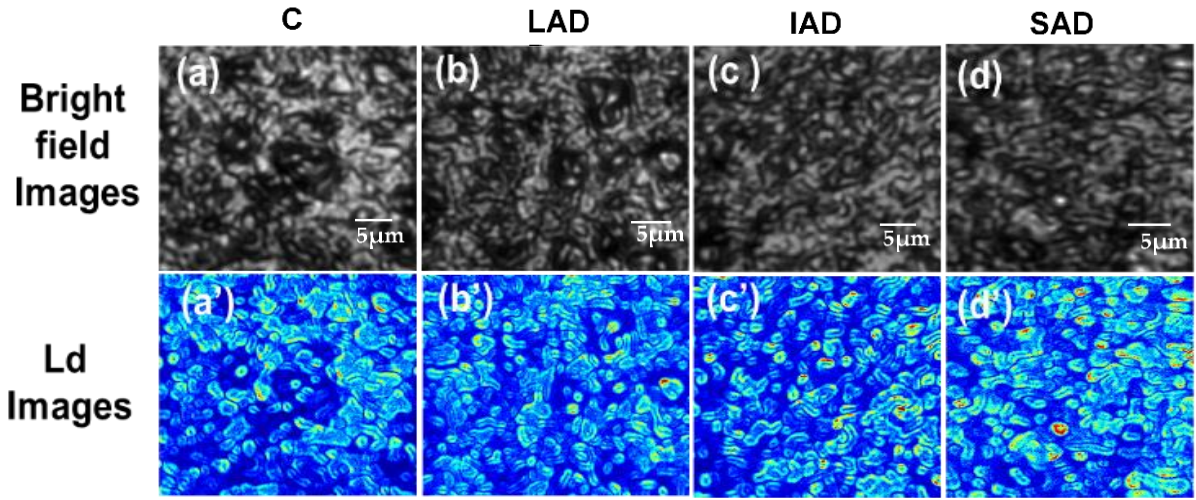

**Figure 1.** $L_{d\text{-}PWS}$ of AD human brain tissue representative samples. (**a**–**d**) Representative brightfield images of healthy tissue and different stages of AD: control (C), low AD (LAD), intermediate AD (IAD), and severe AD (SAD). (**a'**–**d'**) Corresponding $L_d$ images.

We performed a statistical analysis of the ensemble average and standard deviation (std) of the $L_{d\text{-}PWS}$ values, as shown. Figure 2a,b represent the avg and std of the $L_{d\text{-}PWS}$ values of the control and different stages of AD as bar graphs. The error bars in the graphs show the standard error of the mean of the values. The charts show a statistically significant increase in the avg and std ($L_{d\text{-}PWS}$) values for the different stages of AD in comparison to their control. Specifically, the average $<L_{d\text{-}PWS}>$ increased by 6% in LAD, 23% in IAD, and 61% in SAD relative to the control. Similarly, there was an increase in the std ($L_{d\text{-}PWS}$) by 4.2% in LAD, 29% in IAD, and 72% in SAD relative to the control C. These significant increases in structural disorder strength can be attributed to molecular spatial structural changes in the human brain (like DNA methylation and proteomic alterations) as AD progresses.

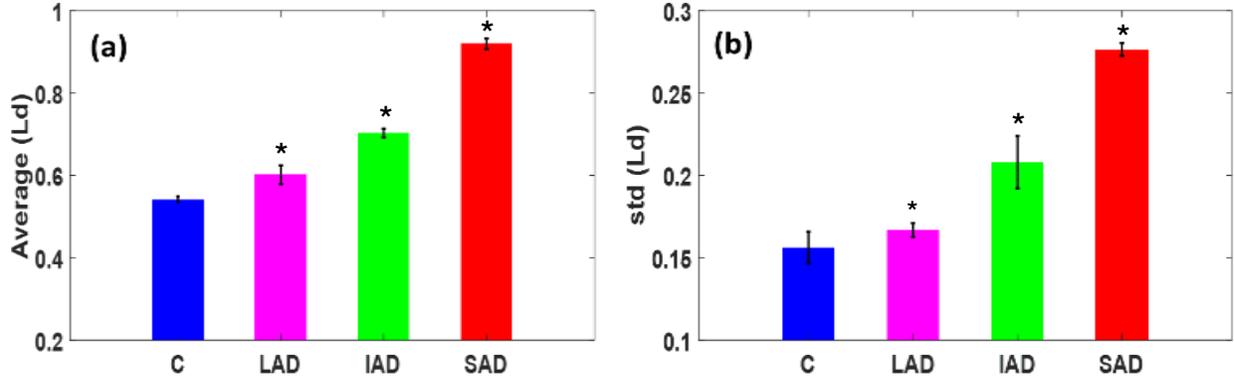

**Figure 2.** PWS analysis of human brain tissue. (**a**,**b**) Bar graphs of the average and std of $L_d$ of the control C and samples of different stages of AD: LAD, IAD, and SAD. The results show an average increase of 6% in LAD, 23% in AD, and 61% in SAD relative to the control C, while there is an increase in the std ($L_{d\text{-}PWS}$) of 4.2% in LAD, 29% in IAD, and 72% in SAD relative to the control. (Student's *t*-test, * *p*-values < 0.05 for each case relative to the control, $N = 10$).

AD has been found to affect the hippocampus and cells at the nanoscale to submicron scale, leading to brain dysfunction. The PWS technique has made it possible to quantify structural alterations in human brain tissues related to AD at the nanoscale level for early signs of AD development and progression. This helps us develop better treatment options, as we better understand structural changes that occur in AD from early to late stages.

*2.2. Confocal IPR Analysis of DNA/Chromatin*

Using the IPR approach discussed in the experimental section, we analyzed the effects of AD on DNA/chromatin spatial mass density fluctuations $d\rho_{ms}$ of nuclei in brain cells embedded in tissues by computing and comparing IPR values with AD progression. We assessed at least five confocal micrographs of each nucleus for each cell. They were evaluated independently at a length scale of *L = 165 nm*. $L_{d\text{-}IPR}$ was computed in terms of IPR for each cell nucleus across the length scale *L*. The average (avg) and standard deviation (std) of the IPR values were calculated by averaging the IPR values of all the tissues at the same scale in every category. These *avg* ($\langle IPR \rangle$) and std ($\langle IPR \rangle$) values of the nuclei were compared between the AD and control samples.

We present the confocal images of the DNA/chromatin nuclei, which were DAPI stained for control C and AD, in Figure 3a,b, along with their respective IPR images shown in Figure 3a',b', captured at a scale of *L × L* (165 nm65 nm). The IPR images can be distinguished easily from their respective confocal images. Structural DNA abnormalities in the cell nuclei are shown clearly, with more red and yellow spots in the

IPR images reflecting more molecular mass density fluctuations. On the contrary, the blue color represents minor mass density fluctuations for each pixel.

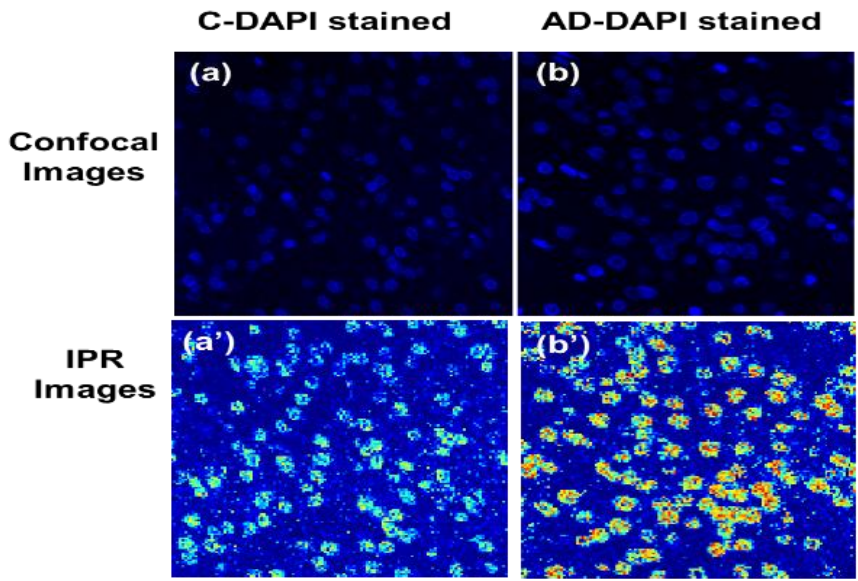

**Figure 3.** DNA molecular-specific structural disorder visualized using confocal-IPR imaging of $L_{d\text{-}IPR}$. (**a**) DAPI-stained confocal image of the control human brain tissue; (**b**) AD human brain tissues that are DAPI stained, targeting DNA/chromatin; (**a',b'**) corresponding $L_{d\text{-}IPR}$ images of the control C DAPI-stained brain tissues and AD DAPI-stained brain tissues, respectively.

Bar graphs were employed to analyze the ensemble of IPR values at a sample length of 165 nm, as shown in Figure 4. The bar graphs depict the avg $(\langle IPR \rangle)$ and std $(\langle IPR \rangle)$ for the DNA mass density variance utilizing the confocal-IPR technique on brain tissues from Alzheimer's patients. When analyzed statistically in Figure 4, it was found that there was an increase in the $L_{d\text{-}IPR}$ of the AD brain tissues relative to the control. There was an average increase of 43% and 50%, respectively, as well as an increase in the standard deviation of the IPR values for the AD in the DNA molecules compared to the control, as shown in Figure 4a,b.

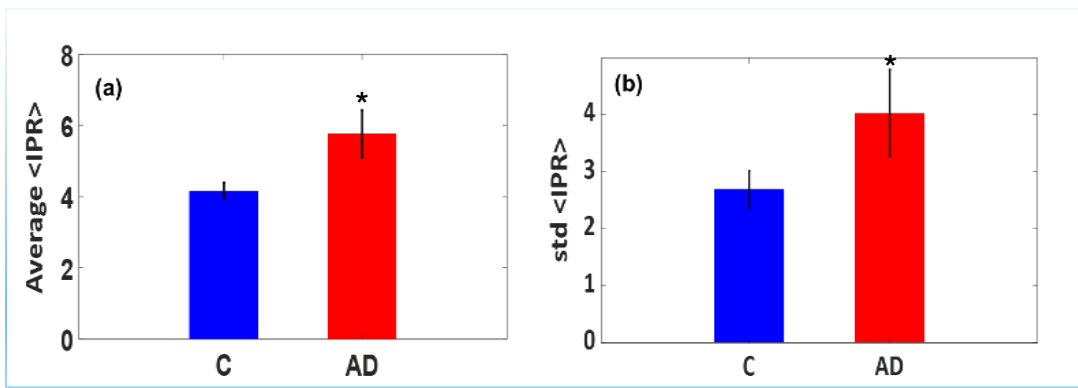

**Figure 4.** $\langle IPR \rangle$ values of DAPI-stained control (C) and AD brain tissue samples. (**a**) Average $\langle IPR \rangle$ value of DAPI-stained C and AD brain tissue samples (**b**) Std $\langle IPR \rangle$ value of DAPI-stained C and AD brain tissue samples. The average and std of $\langle IPR \rangle \equiv L_{d-IPR}$ AD DAPI-stained brain tissue values are higher than the controls (C). The average value of $L_{d\text{-}IPR}$ increased by 50% relative to the control, and its std ($L_{d\text{-}IPR}$) increased by 43% relative to the control C. According to the above bar plots, the AD samples exhibited more disorder in the spatial structural of the DNA in the nuclei of brain cells, leading to more structural damage and abnormalities. (Student's *t*-test, * *p*-value < 0.05 relative to control C, *N* = 10).

This increase in IPR value is related to structural abnormalities, indicating an increase in the DNA spatial molecular mass density fluctuations, $d\rho_{ms}$, or higher molecular fluctuations in the DNA of brain cells. This increase in $d\rho_{ms}$ can be caused by the rearrangement of macromolecules triggered by the deposition of Aβ protein and DNA damage in the brain at the nanoscale.

*2.3. Increased Amyloid Beta Deposition in the Brains of AD Patients*

The accumulation of Aβ protein characterizes AD. To delve deeper into Aβ deposition among AD and non-AD subjects, we employed immunofluorescence staining in hippocampus sections of AD and non-AD brains utilizing Aβ-specific antibodies. There was significant deposition of Aβ protein in the hippocampus sections of the brains of AD patients relative to non-AD patients, as shown in Figure 5.

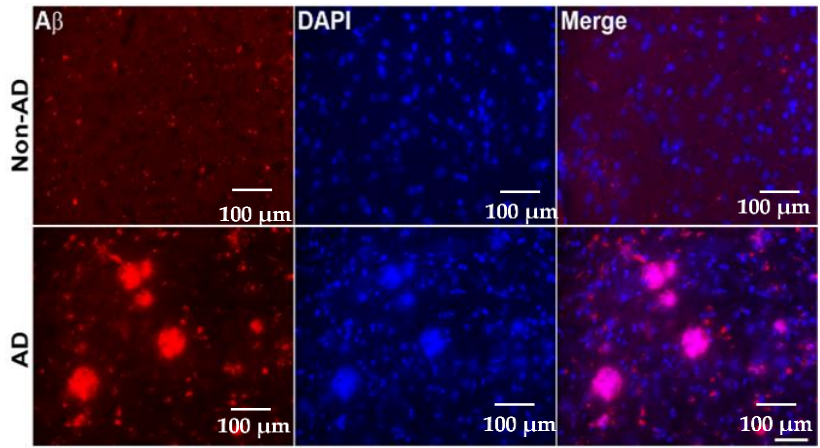

**Figure 5.** Increased amyloid beta (Aβ) deposition in the brains of AD patients. Representative photomicrographs show the profound expression of Aβ deposition (red staining) in AD patients' hippocampi compared to non-AD subjects. Compared to the human non-AD brain, the AD brain showed more immunostained plaques (red), whereas negligible

expression was found in the non-AD brain sections. Microscopic analysis indicates that the fully developed plaques in the brain were amorphous and granular. Scale bar = 100 μm.

*2.4. Increased DNA Double-Strand Breaks (DDSBs) in the Brains of AD Patients*

Next, we investigated if AD is linked to an increase in DDSBs. We performed the ELISA method as discussed in the experimental section. There was a significant increase ($p$-value < 0.05) in the levels of DDSBs in the hippocampi of the AD patients compared to the healthy controls. These results align with prior studies demonstrating increased DNA breaks in AD [28,49,50]. Hence, we found that DDSB accumulation is increased significantly when there are neurodegeneration conditions, as presented in Figure 6.

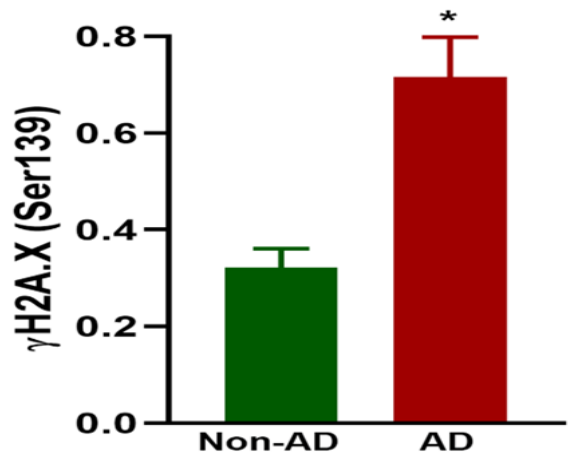

**Figure 6.** Increased DNA double-strand breaks (DDSBs) in the hippocampi of AD patients. The ELISA method was used to assess the protein levels of γ-H2A.X (Ser139) in the hippocampi of AD and non-AD brains. The bar graphs show a net increase of 131% with the progress of AD. Data were analyzed using a two-tailed *t*-test between groups. The values are expressed as means ± SEM. * $p$ < 0.05 (N = 10/group).

**3. Discussion**

Fragmented reports provided in previous studies provide evidence of microscopic structural changes in several brain regions during the asymptomatic stages of AD based on brain biopsies or large-volume imaging such as MRI, X-ray, and CT scans. However, the limiting factors of studying nano-structural changes using similar techniques have been described in previous works [20,23]. This paper includes detailed reports of subcellular alterations in brain tissue in AD, ranging from LAD to SAD. Using the dual spectroscopy techniques of PWS and IPR, we measured and quantified changes at the brain's nano to submicron scale at various stages in AD patients. We focused our study on the different stages to represent the levels of AD progression in human brain tissue, so the measurements are expected to be clinically helpful. Using PWS, the nano-structural alterations in $\rho$;—that is, fluctuations in $d\rho$—increase in the brain tissues of AD patients for an extended period before the symptoms appear. To corroborate these results, we investigated the spatial structural disorder properties of DNA/chromatin using a new photonic technique, IPR, which has proven helpful for probing chromatin/DNA's spatial structural alteration properties in studying the degree of disorder strength properties [52]. The PWS and IPR results show a statistically significant increase in the avg and std of $L_{d-PWS}$ and $L_{d-IPR}$ for AD in human brain tissues at different stages compared to the control. These results indicate a higher degree of spatial molecular structural disorder in DNA, which may be caused DNA damage—a previously undefined parameter in the context of AD—which links structural changes at the nanoscale level to AD onset and progression.

We further performed immunofluorescence staining and ELISAs to examine Aβ deposition and the damage to the DNA, respectively. The results showed that there is ample deposition of Aβ and DNA breaks in the brains of AD patients. The results of these cytological experiments, namely DNA damage, also directly support the optical measure of the DNA damage obtained in the PWS experiment. Accumulation of DNA damage has been linked to several age-related neurodegenerative diseases, including AD [53,54]. In AD, the accumulation of DDSBs is considered an early feature, suggesting that they may act as an initiating lesion of neurotoxicity [55,56]. These observations indicate that the accumulation of DDSBs may degrade chromatin integrity in neurons, ultimately activating deleterious signaling pathways that precede neurodegeneration. Therefore, we can conclude that the accumulation of DNA breaks and structural changes observed via PWS and IPR methods is an essential early part of the pathway toward damage and degeneration in AD.

Lastly, several nanoscale changes in molecular and cellular structures occur in the brains of people with AD because of the damage and death of neurons throughout the brain, as well as the possible breakdown of connections between networks of neurons caused by abnormal accumulations of tau and beta-amyloid proteins inside and between neurons, leading to increasing structural abnormalities and early mass density fluctuations.

## 4. Materials and Methods

### 4.1. Preparation of Brain Tissue Samples

Michigan Brain Bank, Ann Arbor, MI, USA provided samples of brain tissues from AD and cognitively healthy patients. This study received approval from the University of Tennessee Health Science Center Institutional Review Board (IRB #20-07595-NHSR; Exempt Application 874552), Memphis, TN, USA. All standard ethical procedures were followed during this study, including personal protection and safety procedures while handling human tissue samples. Samples collected from AD patients were categorized into different stages by pathologists, which included Low Alzheimer's Disease (LAD), Intermediate Alzheimer's Disease (IAD), and Severe Alzheimer's Disease (SAD).

*4.1.1. Sample Preparation for PWS*

The human tissue samples were embedded in paraffin following a standard protocol, sectioned into 5 μm slices using a microtome, and mounted on glass slides for PWS measurements.

*4.1.2. Sample Preparation for IPR Study Using Confocal Microscopy*

A nuclear dye, DAPI is used to probe the molecular structure of DNA in cell nuclei because it helps to recognize DNA and chromatin structures. Initially, tissue plating was performed on sterile glass slides. Later, the slides were rinsed for 5 min at room temperature in phosphate-buffered saline (PBS) containing 2–4% paraformaldehyde. This cycle was repeated three times before proceeding to the next step. In the next step, the slides were stained using Prolong Diamond antifade mount containing a DAPI stain to label the DNA a blue color. The DAPI in the mountant helped to distinguish the nuclei from the rest of the cell when imaged using fluorescence microscopy.

*4.1.3. Sample Preparation for DNA Damage Analysis*

To investigate damage in the DNA of AD patients, frozen post-mortem brain tissues were collected from AD and non-AD patients from the hippocampus region of the brain. As described in [57], first, the tissues from both groups were treated with γH2A.X (Ser139) Sandwich ELISA. Following the instructions for the PathScan® Phospho-Histone H2A.X (Ser139) Sandwich ELISA kit ( Cell Signaling Technology, Inc. Danvers, MA, USA) we were able to measure and quantify the endogenous levels of γH2A.X (Ser139). The tissues were dissolved in a buffer solution containing a Halt™ protease and phosphatase inhibitor cocktail. These tissues were pipetted into the wells of a coated 96-well plate and incubated at 4 °C overnight. The plates were washed extensively after incubation, and an antibody phosphor-histone H2A.X (Ser139) was added to each well of the plate to detect the captured phosphor-histone H2A.X protein. Another horseradish peroxidase (HRP)-conjugated affinity purified anti-mouse antibody was added to each well to visualize the bound detection antibody. After washing, a substrate solution was added to each well to develop the color. The color formed was proportional to the amount of bound protein in each well. Later, to terminate the reaction, a stop solution was added. A microtiter plate reader (SpectraMax M2e, Molecular Devices LLC, San Jose, CA, USA) was used to check the absorbance at a 450 nm wavelength.

*4.1.4. Sample Preparation for Amyloid Beta (Aβ) Deposition*

Sections from the hippocampus regions were collected and stained using immunofluorescent antibodies, described in detail in [58,59]. Firstly, for the retrieval of antigens, the sections were pretreated with 10 mM citric acid buffer (pH 6.0) followed by 3 PBS washes. Later, the sections were incubated in 5% bovine serum albumin (BSA #A7906; Sigma Aldrich, St. Louis, MO, USA) for 1 h. The sections were exposed to a primary antibody for Aβ (BAM-10; Sigma Aldrich, St. Louis, MO, USA) overnight at 4 °C followed by 3 PBS washes, then they were treated with Alexa Fluor 555 secondary antibody (in vitro) at room temperature for 1 h. All of the sections were cleansed before mounting with Vectashield® (Vector Laboratories Inc., Newark, CA, USA) mounting medium containing DAPI (H-1200, Vector Labs, Newark, CA, USA). Images of the sections were taken at a 40× magnification using a fluorescence microscope.

*4.2. Partial Wave Spectroscopy*

*4.2.1. Optical Setup*

A brief discussion of the setup is described in this report, and a detailed one can be found in our previous works [23,60,61]. Due to the multiple scattering and reflections from the sample and its surface, the backscattered signal $R(x,y,\lambda)$ produced at different wavelengths is focused on the system for imaging and analysis. The PWS setup measures the $L_{d\text{-}PWS}$ of the sample at the specific spatial position (x,y) spectra of the reflecting wavelength ($\lambda$). The PWS setup, described in detail in [24,62,63], consists of a broadband light from a xenon lamp (150 W) (Newport Corporation, Irvine, CA, USA). The light is collimated using a 4f combination of lenses and irises for spatial filtering. The collimated beam is directed onto the sample using a right-angled prism (BRP) and a 50:50 dichroic mirror that focuses the beam via a 40× objective lens (NA = 0.65). The sample was kept on a motorized stage (X–Y axis: 40 nm; Z axis: 100 nm; Zaber Technologies, Vancouver, BC, Canada) for proper focusing. The backscattered light was focused into the CCD for image recording in the visible spectrum (450–700 nm) through a liquid crystal tunable filter (Kurios LCTF) (Thorlabs, Newton, NJ, USA) using a plano-convex lens (f = 10 cm). The synchronized controller allowed the CCD and LCTF to capture the incoming signal in the CCD with an increase in wavelength of 1 nm.

*4.2.2. Calculation of Structural Disorder Strength ($L_{d\text{-}PWS}$)*

The fluctuating component of the backscatter signal was extracted, and a low-pass Butterworth filter was applied at a stable frequency to remove high-frequency signal noise. The backscattered intensity $I(x,y,\lambda)$ is a function of two spatial coordinates. X and Y represent the pixel location in the sample plane, and $\lambda$ is the wavelength. Moreover, to deal with the variations in intensity in the lamp spectrum, a low-order polynomial *Ip($\lambda$)* was fit to *I($\lambda$)*, resulting in the spectrum *R($\lambda$) = I($\lambda$) − Ip($\lambda$)* [62].

The theory of mesoscopic light transport was applied to $R(\lambda)$ in 1D disorder, which is also applicable to electron and dielectric media [63]. It has also been applied to electron and light transport in dielectric media [39]. Cells/tissues can also be considered weakly disordered media. The incoming backscattered light from the sample can be divided into many 1D channels, with statistical information from $dn$ obtained through 1D mesoscopic light transport theory [64,65]. Based on this theory, it has been shown that $L_{d\text{-}PWS}$ can be calculated using the RMS value of the reflection intensity $<R>_{rms}$ and the spectral auto-correlation decay of the reflection intensity ratio $CC(\Delta k)$ [30,63,66,67]. $L_{d\text{-}PWS}$ can be found for any fixed pixel $(x,y)$ and $n_0$ (assume 1.38) as follows:

$$L_{d-PWS} = \left| \frac{Bn_0^2 \langle R \rangle_{rms}}{2k^2} \frac{(\Delta k)^2}{-\ln(\langle C(\Delta k)\rangle)} \right| \tag{1}$$

where $B$ is a calibration constant, $k$ is the wavenumber, and $C(\Delta k)$ is the autocorrelation function, defined as follows:

$$\langle C(\Delta k) \rangle = \frac{\langle R(k)R(k+\Delta k) \rangle}{\langle R(k)R(k) \rangle} \tag{2}$$

$(\Delta k)^2/ln(<C(\Delta k)>)$ is calculated by plotting $-ln(<C(\Delta k)>)$ vs $(\Delta k)^2$ and performing a linear fit. Equation (1) gives us a map of $L_{d\text{-}PWS}$ for each pixel. Furthermore, $L_{d-PWS} = \langle \Delta n^2 \rangle \times l_c$, where $\langle \Delta n^2 \rangle$ is the refractive index variance and $l_c$ is the correlation length of intracellular refractive index fluctuations. The disorder strength is used to evaluate the changes in intracellular density inside the sample [68]. The average and standard deviation of $L_{d\text{-}PWS}$ were computed to analyze the sample's structural alterations.

*4.3. Confocal IPR Technique*

*4.3.1. Confocal Imaging*

A Zeiss710 confocal microscope (Carl Zeiss Microscopy, Jena, Germany) was used to capture 4–8 micrographs above and below a cell nucleus's central plane in Z-stack mode. Selection of the micrographs depended on the most change in the stack based on acquisition of the best coverage of the nuclear area. Later, each cell type was identified based on several images to analyze their components such as DNA and chromatin. Software including ImageJ v1.54c(National Institutes of Health, USA) and MATLAB were applied to combine and process the images. The tissues of the different groups were captured and processed for calculations.

*4.3.2. Inverse Participation Ratio and Analysis of Structural Disorder Strength $L_d$*

The IPR approach is mainly used to study weakly disordered optical media and measure structural disorder by statistically studying their light localization characteristics. Moreover, light localization inside the system can be measured by finding the IPR value of an optical system's eigenfunction, which depends on the interference of light waves [69–71].

IPR is calculated from an eigenfunction defined as $IPR = \int |E(r)|^4 \, dr$, where E is an eigenfunction of the Hamiltonian. It quantifies the degree of spatial localization and further helps us to measure the system's light localization strength, allowing us to determine the degree of structural disorder ($L_{d\text{-}IPR}$) inside the optical lattice.

The refractive index ($n(x,y)$) inside the biological samples is proportional to intracellular ingredients' local mass density distributions. Intensity coming from each sample voxel $(x,y)$ can be quantified as $I(x,y) = I_0 + dI(x,y)$, where $I_0$ is the average intensity. This can be further connected to the refractive index as $n(x,y) = n_0 + dn(x,y) = \rho_{ms0} + \beta \times \rho ms(x,y)$, where $n_0$ is the average refractive index of the sample at $(x,y)$ and $dn(x,y)$ is the refractive index fluctuations [72–74]. This shows us that intensity variations can be used to correlate the fluctuations in refractive index with the local mass density: $n(x,y) \propto M(x,y) \propto I(x,y)$. Using this property, we can construct a refractive index lattice—a disordered optical lattice—by capturing confocal micrographs [35]. Optical potential can be defined as follows:

$$\varepsilon_i(x,y) = \frac{dn(x,y)}{n_o} \propto \frac{dI(x,y)}{I_0} \qquad (3)$$

It is defined at the *i*-th lattice point within the voxel around $(x,y)$.

The Hamiltonian matrix can be obtained by applying Anderson's tight-binding model (TBM), given as $H = \sum \varepsilon_i |i><i|i + t\sum_{\langle ij \rangle} (|i><j| + |j><i|)$, where $|i>$ and $|j>$ are optical eigenvectors at *i*-th and *j*-th lattices, $<ij>$ are the nearest neighbors, and $\varepsilon_i$ is the optical potential energy of the *i*-th lattice site belonging to the pixel position $(x,y)$ [20,31,50]. The mean IPR can be calculated as follows:

$$\langle IPR \rangle_{L \times L} = \frac{1}{N} \sum_{i=1}^{N} \int_0^L \int_0^L E_i^4(x,y) dx dy \qquad (4)$$

The above formula is for an optical lattice size of $L \times L$. $E_i$ is the *i*-th eigenfunction of the Hamiltonian, and the total eigenfunctions are $N = (La)^2$, [$La = L/a$ (lattice size), $a = dx = dy$]. Biological cells are considered heterogenous light transparent systems, which characterize their disorders based on two factors: $dn$ and $l_c$ [75]. Combined with these two parameters, it gives disorder strength, $L_{d-IPR} = <\Delta n> \times l_c$ [76]. It has been reported that the average and the standard deviation of the <IPR> value are correlated to the degree of structural disorder, which can be written as follows:

$$Average \langle IPR \rangle \propto L_{d-IPR} = <\Delta n> \times l_c \qquad (5)$$

$$std \langle IPR \rangle \propto L_{d-IPR} = <\Delta n> \times l_c \qquad (6)$$

## 5. Conclusions

In this article, the results of the dual spectroscopic techniques of PWS and IPR show nano- to submicron-scale structural alterations in tissues/cells and chromatin in progressive Alzheimer's disease from the early to late stages. Using PWS, we probed the structural alterations in brain tissue samples, and using the IPR technique, we probed the spatial structural alterations in DNA/chromatin structures in the brain cells of humans with progressive Alzheimer's disease. The PWS results showed that there was an increase in the degree of structural disorder $L_{d\text{-}PWS}$ ($\sim dn^2.lc$) value of 6% in LAD, 23% in IAD, and 61% in SAD relative to the control C, and their std values increased by 4.2% in LAD, 29% in IAD, and 72% in SAD relative to the control C. Furthermore, the IPR results showed that there was an increase in the degree of structural disorder $L_d\text{-}_{IPR}$ ($\sim dn.lc$) in the DNA/chromatin average value, which increased by 50% in AD relative to the control C. Its std values also increased by 43% in SAD relative to the control C. The significance of this study was the use of light scattering experiments to probe nano- to submicron-scale structural alterations, which was supported by DNA damage histopathology/cytology experiments.

Our experimental findings shed new light on AD and its local increases in the degree of $L_{d\text{-}IPR}/L_{d\text{-}PWS}$ via optical probing of the brain's hippocampus region in AD patients. Understanding the nanoscale structural alterations underlying abnormalities in brain tissues due to AD progression may help in detailed clinical probing for early AD detection and possible treatment. In this study, we only examined the structure changes in the hippocampal region. Therefore, further studies are needed to explore the structural changes in different brain regions and their correlations with AD-related neuropathology.

**Author Contributions:** PP conceptualized the project; P.P. and M.M.K. designed the experiments; F.A., I.A., D.S., S.K., and H.S. performed the experiments; F.A., I.A., D.S., H.S., M.M.K., and P.P. performed the data analysis; F.A., I.A., D.S., M.M.K., and P.P. wrote the first draft, and all authors participated in finalizing the paper. P.P. oversaw the project. All authors have read and agreed to the published version of the manuscript.

**Funding:** This work was partially supported by National Institutes of Health (NIH) grant numbers R21 CA260147 to PP, as well as Alzheimer's Association Award AARG-NTF-22-972518, and Department of Defense Award Number HT9425-23-1-0043 to MMK.

**Institutional Review Board Statement:** This study received approval from the University of Tennessee Health Science Center Institutional Review Board (IRB #20-07595-NHSR; exempt application 874552), Memphis, TN, USA. All standard ethical procedures were followed during this study, including personal protection and safety procedures while handling human tissue.

**Informed Consent Statement:** Human samples were obtained from the Michigan Brain Bank, MI, USA.. Patient identities were not disclosed by the MBB, and this study was approved by IRB #20-07595-NHSR; exempt application 874552.

**Data Availability Statement:** The data may be available upon request to the corresponding author, PP.

**Acknowledgments:** We thank the NIH and DOD for their financial support. We also acknowledge the imaging centers of Mississippi State and UTHSC for confocal imaging.

**Conflicts of Interest:** The authors declare no conflicts of interest.